\documentclass[runningheads]{llncs}
\usepackage{mdframed}
\usepackage{amsfonts}
\usepackage{graphicx,algorithm,algpseudocode}
\usepackage{xcolor}
\usepackage{tabularx}

\newcounter{protocol}

\newcommand{\oded}[1]{\textcolor{black}{#1}}

\begin{document}
\title{JugglingSwap: Scriptless Atomic Cross-Chain Swaps}
%
%
\author{Omer Shlomovits, Oded Leiba}
%
%
\institute{KZen Research}
\maketitle              
\begin{abstract}
The blockchain space is changing constantly. New chains are
being implemented \oded{frequently} with different use cases in mind. As more
and more types of crypto assets are getting real world value there is an
increasing need for blockchain interoperability. Exchange services today
are  still  dominated  by  central  parties  which  require  custody  of  funds. This trust imposes costs and security risks as frequent breaches testify.\\
\oded{\textit{Atomic cross-chain swaps} (ACCS)  allow  mutual  distrusting  parties  to securely exchange crypto assets in a peer-to-peer manner while preserving self-custody.
Fundamental ACCS protocols \cite{nolan,accs} leveraged the scripting capabilities of blockchains to conditionalize the transfer of funds between trading parties.
Recent work \cite{poelstra2017scriptless,malavolta2019anonymous} showed that such protocols can be realized in a scriptless setting. This has many benefits to blockchains throughput, efficiency of swap protocols and also to fungibility and privacy. The proposed protocols are limited to blockchains supporting either Schnorr or ECDSA signatures that are sharing the same elliptic curve parameters.\\
In this work we present \textit{JugglingSwap}, a scriptless atomic cross-chain swap protocol with a higher degree of interoperability. We weaken the assumptions about blockchains that can be included in the ACCS protocol, and only require that (1) a threshold variant exists to the underlying digital signature scheme and (2) it is based on the elliptic curve discrete logarithm problem (ECDLP).} 
The fair exchange is achieved by a gradual release of secrets. To achieve this we \oded{use a new building block we call \textit{Juggling}}: a public key verifiable encryption \cite{cs03} scheme to transfer segments of secret shares between parties, which can also be of separate interest.
Juggling is then tailored to a specific private key management system design with threshold signatures security.\\

\end{abstract}
\section{Introduction}
\subsection{Background}
The problem of fair exchange is how two mutually distrusting parties can jointly
exchange digital assets such that each party receives the other party's
input or neither does. There exist a variety of fair exchange protocols in the literature, all with their own specifications and system model. We divide fair exchange protocols to three categories: 
\begin{itemize}
\item  Trusted Third Party: fair exchange become\oded{s} trivial if both parties can agree on a mutual trusted third party. The parties will send the trusted party the digital assets. The trusted party will verify correctness and redistribute to the counterparty. To make it a more scalable solution there is a line of research to design \textit{optimistically fair} protocols, such that the trusted third party gets involved only in case of disputes which are assumed to be rare \cite{opt_fair_1,opt_fair_2,opt_fair_3}. 
\item Gradual Release of Secrets: This method achieves \textit{partial fairness} such that one party can get advantage over the other party but this advantage is polynomially bounded. The  idea is to enable each party to release their secret bit by bit in alternation \cite{grad_rel}.
\item Monetary Penalties:  parties are incentivized to complete the protocol fairly, and
if one party receives its output but aborts before the other party does, the
cheating party will have to pay a penalty \cite{escrow,kumaresan15poker,lindell_legal}.
\end{itemize}
Focusing on blockchains, \oded{fair exchange is a precursor for escrow protocols with cryptocurrencies \cite{escrow}, markets, auctions, games \cite{kumaresan15poker} and atomic cross-chain swaps (ACCS).
An ACCS is a task in which multiple parties wish to exchange digital assets from possibly distinct blockchains, in such a manner where all parties receive their desired output from another party, or neither does. In this paper we consider specifically the two-party setting.\\
Blockchains are decentralized by nature, therefore a core challenge is to design protocols without a single point of failure in the form of a trusted third party which takes full custody of users' funds. Early ACCS protocols leveraged the blockchain itself as a decentralized trusted party that can lock and release funds according to pre-programmed rules.
The first successful atomic cross-chain swap construction is credited to TierNolan \cite{nolan}. It leverages Bitcoin-like script for creating conditional transactions by the two exchanging parties: each party locks its funds such that the counterparty can spend them by using the preimage of an agreed hash value. When one party spends a transaction, it reveals the secret to the other party which can use it to spend the other conditional transaction on the other blockchain. There is also a timeout after which the funds can be refunded if not redeemed by the counterparty. Such conditional spending is termed Hashed Timelock Contract (HTLC) and has been specified and extended in \cite{LN} and \cite{bip199}. Herlihy  in \cite{accs} has formalized the theory of such atomic cross-chain swaps. Poelstra \cite{poelstra2017scriptless} has introduced the notion of \textit{scriptless scripts} and described how to achieve atomic swaps without the need of explicit scripting within transactions. Instead, Schnorr signatures \cite{schnorr1991efficient} are used: the spending of a transaction reveals a secret merely by the unlocking signature itself, and only to the counterparty. This has many benefits to blockchains throughput, efficiency of swap protocols and also to fungibility and privacy. Malavolta et al. \cite{malavolta2019anonymous} has extended this technique to ECDSA. The latter two studies have opened the door for scriptless atomic cross-chain swaps for a variety of existing blockchains. However, they are still limited to blockchains which support either Schnorr or ECDSA signatures that share the same elliptic curve parameters, namely - the same group generated from the same fixed generator.}

\subsection{Our Contribution}
Our solution, \oded{the \textit{JugglingSwap} ACCS protocol,} is based on gradual release of secrets. We observe that there are two ways to transfer value in a blockchain: 
\begin{itemize}
\item  A transaction backed by a digital signature. 
\item  A transfer of a secret key.
\end{itemize}

To the best of our knowledge, all existing exchange methods use the first method. \oded{The first method naturally has more constraints to it: either it depends on the scripting capabilities of the underlying blockchain \cite{accs,nolan,bip199} or on specific mechanics of the underlying signature scheme \cite{poelstra2017scriptless,malavolta2019anonymous}}. The second method is purely cryptography-dependent with small variance between blockchains and thus can fit \oded{and be reused} for many chains architectures.\\
The vast majority of blockchains tie funds to key pairs of an elliptic curve group where the elliptic curve discrete logarithm problem (ECDLP) is considered hard. Exchanging between two chains in that case comes down to encrypting the secret key with the other party's public key.
The two problems that arise: 
\begin{enumerate}
\item When one party transfers a secret key to a second party it does not lose knowledge of this secret key. The transfer is like a "copy" whereas in a transaction based swap the transfer is like a "cut": the sending party loses access to the funds. The undesired result is that both sending and receiving parties can use the secret key at the end of the transfer.
\item The exchange is not fair. The first to send the encryption will risk the other party to not complete the protocol.
\end{enumerate}
We address these issues in the following way: first, we add a layer of threshold cryptography for secret key management. The cross-chain interoperability remains and we get better security. We start our protocol assuming that each of the two trading parties, the \textit{owners}, has a joint \textit{two-out-of-two} ($\{2,2\}$) address with the same centralized \textit{provider}.\\
\oded{Second, we introduce and use a new building block we call \textit{Juggling}: a novel verifiable encryption \cite{cs03} construction for transferring segments of a discrete logarithm. It allows a prover to encrypt a secret in parts and prove that each resulting ciphertext is indeed an encryption of a segment of a secret key of some known public key, under some known encryption key.\\
Combining these two building blocks, the ACCS protocol goes as follows:
\begin{enumerate}
\item The two owners and the provider generate two $\{3,3\}$ addresses: one on each of the two blockchains. Each party keeps a local secret key share.
\item Each owner deposits funds into the $\{3,3\}$ address of its source blockchain.
\item The two owners start executing two interleaving Juggling protocols, in order to achieve a \textit{partial fair exchange} of their respective secret shares of the two $\{3,3\}$ addresses.
In other words - they gradually release their secret shares, segment by segment. Eventually both owners hold 2 shares of their counterparty's $\{3,3\}$ address and only 1 share of their own $\{3,3\}$ address. 
\item Each owner co-signs with the provider in order to withdraw the funds from the counterparty's $\{3,3\}$ address to its own wallet.
\end{enumerate}}

During the execution of the Juggling protocol (step 3), in case one party decides to cheat, the maximal advantage is to be one segment ahead. For small enough segment bit size this is acceptable. 
The provider's role can theoretically be played by a trusted execution environment (TEE).\\
\oded{The resulting ACCS protocol is scriptless. The blockchain footprint of our method is between one to two regular transactions on each blockchain. Therefore, it offers similar benefits as the ACCS protocols in \cite{poelstra2017scriptless,malavolta2019anonymous} in terms of improved blockchain throughput and more efficient swaps (cheaper than their script-dependant alternatives), as well as in fungibility and privacy. On top of that, because Juggling is applicable on arbitrary groups, JugglingSwap has weak assumptions on the participating blockchains: it only requires that (1) a threshold variant exists for the underlying signature algorithm and (2) it is based on ECDLP. Therefore, the set of pairs of blockchains supported by our protocol strictly contains the set of possible pairs by prior art, so it offers a higher degree of interoperability.}
Finally, We believe that the Juggling protocol can be of separate interest and used as a building block for other constructions as well.
\subsection{Related Work}
\subsubsection{Gradual Release of Secrets:} The notion of partial fairness, originated in \cite{partial1,partial2}, was proposed as a way to overcome the impossibility result of perfect fairness in a setting without a trusted third party \cite{cleve}. Several protocols, such as \cite{damgard,damgard2}, were built on top of this security notion. Previous works are mainly focused on bit by bit release of either the secret itself or a digital signature from the secret and are not generalizable to elliptic curves DLog. Our protocol is aimed for gradually releasing a ciphertext encrypting an EC-DLog, where the encryption correctness is publicly verified for every segment and segment size is configurable. 
\subsubsection{Crypto Asset Exchange:}
Several types of exchange methods exist, most of them are used in practice today. In centralized-custodial exchanges (e.g. \cite{binance}), the exchange is a trusted third party. Funds are deposited to the exchange platform, trading is done within the platform and finally the funds can be withdrawn to a private wallet. From the moment of the deposit and until the withdrawal the funds are owned by the exchange which at any moment can be hacked, stop working or become malicious and steal all the funds currently held by the platform. \\
Centralized but non-custodial trading systems (e.g. \cite{shapeshift}) are platforms that guarantee security before and after the trade but not in the middle: after funds have been sent to the system by the user and before they are returned, the user must trust the trading platform, which again can be hacked, suffer denial of service (DoS) or become malicious and steal the funds. The presence of such a single point of trust clearly does not match well the decentralized nature of blockchain technology.\\
Another form of exchange is decentralized exchanges (or "DEXes", e.g. \cite{zerox,uniswap}). These are trustless platforms that settle trades on the blockchain. They can suffer from security issues such as frontrunning and transaction reordering \cite{frontrun,daian19flash}. The on-chain trading makes scalability another issue. \oded{Lastly, as they typically utilize smart contracts, they are limited to support only the assets of the specific blockchain the contracts are deployed on.}\\
\textit{Tesseract} \cite{tess} is using Intel SGX enclave as a trusted execution engine to provide a better trusted centralized custodial exchange. The problem with this method is that SGX can suffer from security vulnerabilities as shown recently \cite{sgx}. \textit{Arwen}'s \cite{heilman2020arwen} protocol offers centralized non-custodial approach based on building blocks that are used in the Lightning network protocol \cite{LN}, namely an escrow service with off-chain trading and on-chain settlement. The problems with this protocol are the same as the problems with the Lightning network, for example:  (1) limited capacity, (2) an exchange needs to have many coins tied up to escrow channels, (3) fits only blockchains with specific features such as expiry time transactions and (4) requires a new implementation for every blockchain added.\\

\section{Preliminaries}
Here we provide high level informal definitions for Threshold Signatures, Verifiable Encryption and Bulletproofs range proofs. For a more detailed treatment we refer to the original papers. 
\subsection{Threshold Signatures}
Let ${\cal{S}}$=\textrm{(Key-Gen, Sign, Ver)} be a signature
scheme. Following \cite{dss96} A $\{t, n\}$-threshold signature scheme ${\cal{TS}}$ for ${\cal{S}}$ is a pair of protocols
\textrm{(Thresh-Key-Gen, Thresh-Sign)} for the set of parties ${P_1, . . . , P_n}$.
\textrm{Thresh-Key-Gen} is a distributed key generation protocol used by the players to
jointly generate a pair ${(Q, x)}$ of public/private keys on input a security parameter ${1^\lambda}$
. At the end of the protocol, the private output of party ${P_i}$
is a value ${x_i}$ such
that the values ${(x_1, . . . , x_n)}$ form a $\{t, n\}$-threshold secret sharing of ${x}$. The public
output of the protocol contains the public key ${Q}$. Public/private key pairs ${(Q, x)}$
are produced by \textrm{Thresh-Key-Gen} with the same probability distribution as if
they were generated by the \textrm{Key-Gen} protocol of the regular signature scheme ${\cal{S}}$.
\textrm{Thresh-Sign} is the distributed signature protocol. The private input of ${P_i}$
is the value ${x_i}$. The public inputs consist of a message ${m}$ and the public key ${Q}$.
The output of the protocol is a value ${sig \in }$\textrm{Sign}${(m, x)}$.
The verification algorithm for a threshold signature scheme is, therefore, the
same as in the regular centralized signature scheme ${\cal{S}}$

\begin{definition} We say that a ${(t, n)}$-threshold signature scheme ${\cal{TS}}$ =\textrm{(Thresh-Key-Gen,Thresh-Sign)}
is unforgeable, if no malicious adversary who corrupts at
most ${t-1}$ players can produce, with non-negligible (in ${\lambda}$) probability, the signature
on any new (i.e., previously unsigned) message ${m}$, given the view of the protocol
\textrm{Thresh-Key-Gen} and of the protocol \textrm{Thresh-Sign} on input messages ${m_1, . . . , m_k}$
which the adversary adaptively chose.
\end{definition}
This is analogous to the notion of existential unforgeability under chosen message
attack as defined by \cite{digsig}. Notice that now the
adversary does not just see the signatures of ${k}$ messages adaptively chosen, but
also the internal state of the corrupted players and the public communication of
the protocols.\\
We call the special case of $t=n$, namely a protocol which allows a group of signers to produce a short, joint
signature on a common message where all participants are required to be honest, a \textit{multi-signature}.
\subsection{Verifiable Encryption}
Loosely speaking, verifiable encryption \cite{cs03} for a relation $\cal{R}$ is a protocol that allows a prover to convince
a verifier that a given ciphertext is an encryption under a given public key of a value $\omega$ such that
$(\delta, \omega) \in \cal{R}$ for a given $\delta$.\\
Let (Gen, Enc, Dec) be a public key encryption scheme, and let $(\mathrm{pk}, \mathrm{sk})$ be a key pair.
A verifiable encryption scheme proves that a ciphertext encrypts a plaintext satisfying a certain
relation $\cal{R}$. The relation $\cal{R}$ is defined by a generator algorithm $G'$ which on input $1^{\lambda}$ outputs a
description $\Psi = \Psi[\cal{R}, W$, $\Delta ]$ of a binary relation $\cal{R}$ on $\cal{W} \times $$\Delta$. We require that the sets $\cal{R},\cal{W}$ and
$\Delta$ are easy to recognize (given $\Psi$). For $\delta \in \Delta$, an element $\omega \in \cal{W}$ such that $(\delta, \omega)\in \cal{R}$ is called a witness for $\delta$. The idea is that the encryptor will be given a value $\delta$, a witness $\omega$ for $\delta$, and then encrypts $\omega$ yielding a ciphertext $\psi$. After this, the encryptor may prove to
another party that $\psi$ decrypts to a witness for $\delta$. In carrying out the proof, the encryptor
will of course need to make use of the random coins that were used by the encryption algorithm: we denote by Enc'$(\mathrm{pk}$, m) the pair ($\psi$, $coins$), where $\psi$ is the output of Enc$(\mathrm{pk}, m)$ and $coins$ are
the random coins used by Enc to compute $\psi$.
In such a proof system, the (honest) verifier will output 0 or 1, with 1 signifying “accept.”
We require that the proof system is sound, in the sense that if a verifier accepts a proof, then with overwhelming probability, $\psi$ indeed decrypts to a witness for $\delta$. However,
it is convenient, and adequate for many applications, to take a more relaxed approach: instead
of requiring that $\psi$ decrypts to a witness, we only require that a witness can be easily
reconstructed from the plaintext using some efficient reconstruction algorithm. Such an algorithm
$recon$ takes as input a public key $\mathrm{pk}$, a relation description $\Psi = \Psi[\cal{R}, W$, $\Delta ]$, an element $\delta \in \Delta$, and a
message m $\in \mathrm{M_{pk}}\cup \{reject\}$, and outputs $\omega \in \cal{W}$ $\cup$ $\{reject\}$.\\
\begin{definition}
 A proof system $(\cal{P}, \cal{V})$, together with mutually compatible encryption scheme $(Gen, Enc, Dec)$, relation generator $G'$, and reconstruction algorithm $recon$ , form a verifiable encryption
scheme, if Correctness, Soundness and Special honest-verifier zero knowledge properties hold as defined in \cite{cs03}.
\end{definition}

\subsection{Bulletproofs Range Proofs}
Range proofs are proofs that a secret value, which has been encrypted or committed
to, lies in a certain interval. Range proofs do not leak any information about the secret value, other
than the fact that they lie in the interval. Bulletproofs \cite{bulletproof} \oded{can be used as an efficient instantiation of range proofs} and work on Pedersen commitments \cite{pedersen}. Formally, let group element $V$ be a Pedersen commitment to value $v$ using randomness $r$. The proof system will convince the verifier that $v \in [0,..,2^n-1]$. Bulletproofs are based on Inner Product Argument and maintain perfect completeness, perfect special honest verifier zero-knowledge, and computational witness extended emulation, defined in \cite{bulletproof}. 

\section{The Juggling Protocol}
\subsection{Specification and Definitions}
Let $\mathbb{G}$ be an elliptic curve group of prime order $q$ with base point (generator)
$G$. Party $P_i$ knows discrete log (DLog) $x_i$ such that $Q_i=x_iG$.
 
\begin{definition}
m-Segmentation of $x$ is the division of $x$ to $m$ segments $[x]_k|_{k=1}^{m}$ such that $x = \sum_{k}{f_k[x]_k}$ where $f_k = 2^{(k-1)l}$ for segment length in bits $l$.
\end{definition}
\begin{definition} \label{proof_of_correct_k}
We say that a proof is a \textbf{proof of correct encrypted k-segment} if it is a publicly verifiable zero knowledge proof for encryption scheme $\{Gen, Enc,$ $ Dec\}$, public key $Y$, value $\delta$, relation $\mathcal{R}$, ciphertext $c_k$ and parameters $k,m$ such that $c_k=Enc(Y,[x]_k)$ is the encryption of $[x]_k$ using public key $Y$, where $[x]_k$ is the $k$'th segment of m-Segmentation of a witness $\omega=x$ such that $(\delta,\omega) \in \mathcal{R}$.
\end{definition}
\begin{definition}
 A proof system $(\cal{P}, \cal{V})$, together with mutually compatible encryption scheme $(Gen, Enc, Dec)$ and parameter $m$ is a \textbf{Segmented-Verifiable Encryption (S-VE)} if (a) it is a VE for witness $\omega$ and (b) the witness can be m-Segmented such that for each segment $k$ the prover can prove correct encrypted k-segment for the same witness $\omega$. 
\end{definition}
\begin{definition} \label{juggling_def}
Juggling protocol is S-VE implementation where the witness is an elliptic curve DLog and segment encryptions and proofs are being released serially. 
\end{definition}
It is assumed that there exists a PKI such that all public keys are registered. If it is not the case, all participants will add a setup phase for key pairs generation and registration. 
\subsection{Auxiliary Proofs} \label{aux_proofs}
All proofs are for public EC parameters $\mathbb{G},q,G$.
\subsubsection{Proof of EC-DDH membership:} \label{proof_of_membership}
This is a proof of membership that a tuple is an Elliptic Curve Decisional Diffie Hellman (DDH) tuple: $(G,xG,yG,xyG)$.
This proof is an adaptation of the proof in \cite{cp93} for elliptic curves. 
Proof for the following relation: the witness is $\omega=x$, the statement is $\delta=(G_1, H_1, G_2, H_2)$. The relation $\mathcal{R}$ outputs $1$ if ${H_1 = xG_1}$ and ${H_2 = xG_2}$. 

The protocol works as follows:
\begin{enumerate}
\item The prover sends $a_1 = \alpha G_1$ and $a_2 = \alpha G_2$ to the verifier with $\alpha \in_R Z_q$
\item The verifier sends a random challenge $e \in_R Z_q$
\item The prover responds with $r = \alpha - xe $
\item The verifier checks that $a_1 = rG_1 + eH_1$ and $a_2 = rG_2 + eH_2$
\end{enumerate} 
See \cite{cp93} for the formal proof.
\subsubsection{Proof of correct encryption:} \label{proof_of_enc_dlog}
This is a proof of knowledge that a pair of group elements $\{D,E\}$ form a valid homomorphic ElGamal encryption ("in the exponent") \cite{elgamal} using public key $Y$. Specifically, the witness is $\omega=(x,r)$, the statement is $\delta=(G,Y,D,E)$. The relation $\mathcal{R}$ outputs $1$ if $D=xG+rY$, $E=rG$. 

\begin{enumerate}
\item The prover chooses $s_1,s_2$, computes $A_1=s_1G,A_2=s_2Y, A_3 = s_2G$ and sends to the verifier $T = A_1 + A_2, A_3$.
\item The verifier picks a challenge $e$.
\item The prover computes $z_1=s_1+ex$ and $z_2=s_2+er$.
\item The verifier accepts if $z_1G + z_2Y=T+eD$ and $z_2G=A_3+eE$.
\end{enumerate} 

\subsubsection{Proof of correct encryption of DLog:} \label{proof_of_enc}
This is a version of the previous proof where we also know public key $Q$ and want to prove that $\{D,E\}$ is homomorphic ElGamal encryption that encrypts the DLog of $Q$, specifically. 
This is a proof for the following relation: the witness is $\omega =(x,r)$, the statement is $\delta=(G,Y,Q,D,E)$. The relation $\mathcal{R}$ outputs $1$ if $Q=xG$, $D=xG+rY$, $E=rG$. To prove this the prover needs (a) to prove knowledge of the DLog of $Q$ in base $G$ and (b) to prove that $(G,E,Y,D-Q)$ is a DDH tuple. The proof can be optimized by using the same challenge as follows:  

\begin{enumerate}
\item The prover chooses $s_1,s_2$, computes $A_1=s_1G,A_2=s_2Y, A_3 = s_2G$ and sends $A_1,A_2,A_3$ to the verifier.
\item The verifier picks a challenge $e$.
\item The prover computes $z_1=s_1+ex$ and $z_2=s_2+er$.
\item The verifier accepts if $z_1G=A_1+eQ$ and $z_2G=A_3+eE$ and $z_2Y=A_2+e(D-Q)$.
\end{enumerate} 
The protocols can become non-interactive using Fiat-Shamir transform. Both Constructions: proof of correct encryption and proof of correct Encryption of DLog are Sigma protocols \cite{carmit} with straight forward security proofs. We provide the proofs in Appendix A.

\subsection{Construction} \label{construction}
\begin{figure}[htbp]
\centering
  \includegraphics[trim=2cm 7cm 2cm 0, clip, width=\textwidth]{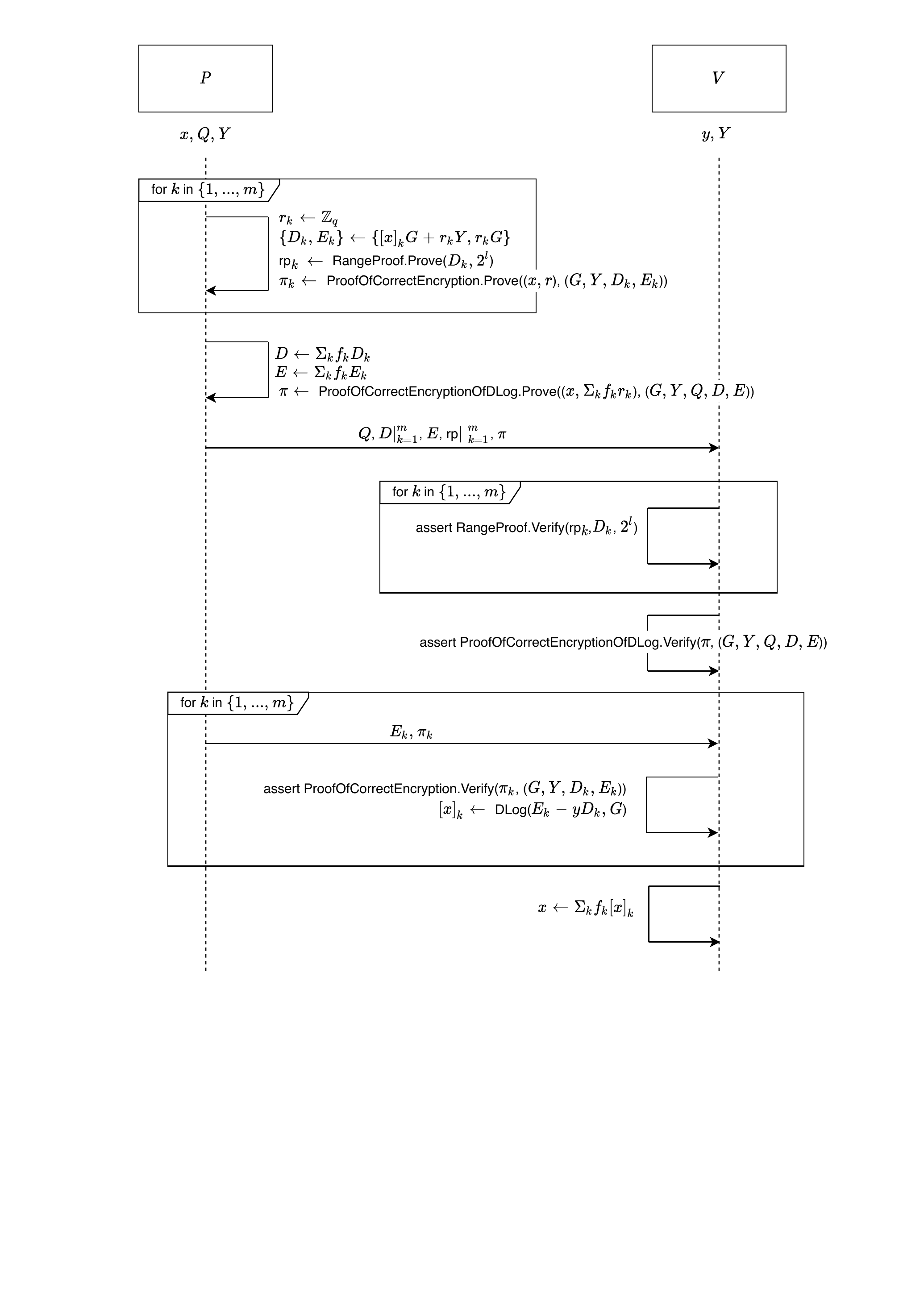}
  \caption{\textbf{The Juggling protocol sketch}.
  Here $P$ is the encryptor/prover, and $V$ is the decryptor/verifier. Proofs and encryptions are being released serially by $P$, where after each iteration they are being verified and decrypted by $V$.\newline
Note: when both parties fulfill both roles (each encrypting its own secret), these iterations can interleave: the first party receiving the k'th segment will then send back its own k'th segment.}
  \label{fig:juggling-sketch}
\end{figure}

We observe that S-VE can be implemented using an encryption scheme with homomorphic properties. We choose homomorphic ElGamal encryption as the base encryption scheme \cite{elgamal}. \oded{Figure \ref{fig:juggling-sketch} shows a sequence diagram of the protocol.}\\


\begin{mdframed}[userdefinedwidth = 12.3cm]
\begin{enumerate}
\item \textbf{Key Generation:} \newline
\newline
For EC parameters $\mathbb{G},q,G$ and security parameter $\lambda$ every party chooses random $y\in\mathbb{Z}_q$ for secret key and computes $Y=yG$ for public key. All public keys are registered.\newline 
\item \textbf{Encryption:} \newline
\newline
 Upon input $(x,Q,Y)$ where $Q,Y$ are public keys and $x$ is a secret key such that $Q=xG$ the encryptor (interchangeably will be called the prover, depends on operation) divides $x$ to $m$ equal segments (last segment can be padded with zeros). The segments $[x]_k|_{k=1}^m$ should be small enough to allow extraction of $[x]_k$ from $[x]_kG$ in polynomial time in the security parameter. 
\begin{enumerate}
\item For every segment $k$: the encryptor computes the homomorphic ElGamal encryption :
$\{D_k,E_k\}=\{[x]_{k} G+r_k Y,\  r_k  G\}$ for random $r_k$. 

\item The encryptor publishes $D_{k}|_{k=1}^{m}$ together with $m$ Bulletproofs (\cite{bulletproof}, non-interactive zero knowledge range proofs) proving that every $D_{k}$ is a Pedersen commitment \cite{pedersen} with value smaller than $2^l$ (*).

\item The encryptor publishes $E = \sum_{k}{f_kE_{k}}$ and plays the prover in \textit{proof of correct encryption of DLog} (\ref{proof_of_enc_dlog}) with witness $(x, \sum_{k}{f_kr_k})$ and statement $(G,Y,Q,D,E)$, where $D = \sum_{k}{f_kD_{k}}$.

\item An encryption for a segment $[x]_k$ will be full once the encryptor publishes $E_k$ of this segment together with a \textit{proof of correct encryption} (\ref{proof_of_enc}). \newline
\end{enumerate} 

\item \textbf{Decryption:} \newline
\newline
 Given a secret key $y$, for every pair $\{D_k,E_k\}$, $[x]_k$ can be decrypted by extracting $[x]_k$ from $D_k-yE_k = [x]_kG$ using algorithm for breaking DLog. After all $m$ segments are decrypted $x=\sum_{k}{f_k[x]_k}$ can be reconstructed.
\end{enumerate}
(*) Except to the MSB segment which is tightly bounded to the $l$ most significant bits of $q$.
\end{mdframed}
\centerline{Juggling Scheme}

 \subsection{Security Analysis}
 In this section we show that the construction is following the specifications.
 \begin{lemma} \label{lemma1}
 If DDH is hard relative to $\mathbb{G}$, then the scheme \{Key Generation, Encryption, Decryption\} is CPA-secure encryption scheme. 
 \end{lemma}
\textit{Proof sketch:} ElGamal encryption in the exponent, like standard ElGamal encryption is CPA-secure under DDH \cite{intro_to_crypto,elgamal}. $\{D_k,E_k\}$ is ElGamal encryption in the exponent with randomness $r_k$. Because the randomness $r_k$ is changed for every encryption $k\in\{1,..,m\}$ the different encryptions are unrelated. 
\begin{lemma} \label{lemma2}
The proof system $\{\mathcal{P},\mathcal{V}\}$ in $Encryption$ protocol together with homomorphic ElGamal encryption scheme is a \textit{proof of correct encrypted k-segment} of DLog of $Q$, with parameters k, m, Y and security parameter $\lambda$.
\end{lemma}

\textit{Proof sketch:} The proof system consists of $2m + 1$ proofs given in a specific order and correspond to gradual release of the ciphertext: 
\begin{enumerate}
\item $m$ Bulletproofs.
\item Proof of correct encryption of DLog.
\item Proof of correct encryption for first segment
\item ...
\item Proof of correct encryption for the $m$'th segment
\end{enumerate}

the order of the last $m$ proofs can be different and it is easy to show that it does not effect security. \\
The security proof shows that only $m+2$ proofs are needed to proof correct encrypted k-segment. Specifically, for segment $k$ there is no use for the $m-1$ proofs of correct encryption for the other segments. \\
\textit{Correctness:}
This part of the proof is where we need range proofs, which are efficiently instantiated by Bulletproofs. To intuitively understand why, let us first describe a possible attack and why bulletproofs can mitigate it. Proof of correct encryption for the $k$'th segment proves to the verifier that $\{D_k,E_k\}$ is homomorphic ElGamal encryption using public key $Y$. The general proof of correct encryption proves to the verifier a statement about the \textit{shifted sum} ($\sum_k{f_k}$) of the $D_k$'s and $E_k$'s. Namely, that $\{D,E\}$ is Homomorphic ElGamal encryption of $x$ under public key $Y$. A simple attack would be to encrypt biased segments such that the shifted sum will remain the same but each segment encryption will be decrypted wrong. W.l.o.g let us change only two segments: for segment $k$ to encrypt the point $v_k = u_kG = [x]_{k}G + b_{k}G = [x]_kG + f_{k}^{-1}G$ and for segment $k+1$ to encrypt the point $v_{k+1} = u_{k+1}G = [x]_{k+1}G + b_{k+1}G =  [x]_{k+1}G - f_{k+1}^{-1}G $. After decryption, value extraction and segment summation the result would remain the same and equal to $x$ but both segment encryptions $\{D_k,E_k\},\{D_{k+1},E_{k+1}\}$ would have extracted to wrong values. In particular the $k$'th segment will be extracted to $[x]_k +  f_{k}^{-1} \neq [x]_k$. The implication of such attack is that if a prover aborts in the middle of the gradual release, then the decryptor will remain with nothing while it was expected that it will have all the already published segments of the secret. We now show how assuming Bulletproof security prevents this specific attack and then generalize it to all possible changes to encrypted segments $[x]_k|_{k=1}^{m}$. In our attack there are four cases: one case where both $u_k,u_{k+1}$ are in range $[0,...,2^l-1]$ and three cases where at least one of $u_k,u_{k+1}$ are not in the range $[0,...,2^l-1]$. For the last three cases it is easy to see that at least one Bulletproof range proof will not verify with high probability. For the first case in order for the proof of correct Encryption of DLog to pass we require the sum $b_kf_k + b_{k+1}f_{k+1} = nq$, to equal a multiple of $q$, the order of the curve. Every option other than $n=0$ and $b_k=b_{k+1}=0$  will result in at least one of $b_k, b_{k+1}$ out of range $[0,...,2^l-1]$ with high probability. \\
We are now ready to generalize this argument:\\
\textbf{Claim 1.} For any $k$ segment ciphertext not in the form $[x]_k +b_k$ with $b_k$ independent of $[x]_k$ verification will pass with $negl(\lambda)$ probability.\\
\textit{Proof:} Assuming correctness of $m$ bulletproofs, proof of correct encryption of DLog and proof of correct encryption of $k$ segment, if the $k$'th ciphertext is of different form it means that at least one other segment must encrypt an additional compensating element in the $k$'th segment to cancel the change. This additional element would have cause the range proof for this segment to fail with probability $1-negl(\lambda)$. \\
\textbf{Claim 2.} For any $k$ segment ciphertext in the form $[x]_k +b_k$ , if $b_k \neq 0$  verification will pass with $negl(\lambda)$ probability.\\
\textit{Proof:} The following must hold: $\sum_k{b_kf_k} = nq$ If $b_k \neq 0$ for some $k$ at least one of the $b_k$'s will lead to $u_k = [x]_k + b_k > 2^l$. Assume to the contrary that for all $k$: $b_k < 2^l - [x]_k$ and $q=\sum_k{b_kf_k}$. Combining this equations we get $x < \sum_k{2^lf_k} - q$. We assume for simplicity that the most significant segment of $q$ is all-ones. Usually this will restrict the secret key to a very small number such that for random secret key there is negligible probability that the equation will hold. If the secret key is not chosen randomly or if we want to make the probability lower we can simply use secret keys with most significant bit fixed and equal to $0$ (This will not effect security, see discussion in \cite{lindell17} on using secret key $x<q/3$). This can be also proven in Encryption time by providing a range proof on $\{D,E\}$. Now the most significant segment can be range proofed to be in the range $[0,...,2^{l-1}-1]$ such that $2^{ml-1} + \sum_{k=1}^{m-1}{2^lf_k} - q$ is negative and therefore assuming bulletproof security, this is possible with only $negl(\lambda)$ probability.  \\
\newline
\textit{Soundness:} Soundness comes from the soundness of proof of correct encryption of DLog. There exist algorithm $\mathcal{A}$ that given $Q$ and two transcripts of this proof with two different challenges $e \neq e'$ outputs $\omega=x$ such that $Q=xG$. This is enough because algorithm $\mathcal{A}'$ for soundness of the proof of correct encrypted k-segment of DLog $Q$ can use $\mathcal{A}$ and output the $k$'th segment of the result.\\
\newline
\textit{Special honest verifier ZK:} All three types of proofs: bulletproof, proof of correct encryption of DLog and proof of correct encryption have the special HVZK security property, meaning there exist a simulator $S$ that can output computational indistinguishable distribution by calling the special HVZK simulators of the underlying relevant proofs.

\begin{theorem}
Assuming DDH is hard in $\mathbb{G}$ construction \ref{construction} is a Juggling protocol according to definition \ref{juggling_def} in the random oracle model.
\end{theorem}
\textit{Proof sketch:} The proofs of correct encryption combined form a VE of witness $x$ and DLog relation $(Q,x)\in \mathcal{R}$ if $Q=xG$. The prover cannot change witness from the range proof to the \textit{proof of correct encryption of DLog} since otherwise the prover would have broken the binding property of $D_k$ as a commitment, which is equivalent to finding the DLog of $Y$ which assume to be hard. Same is true for changing witness from proof of correct encryption of DLog to the \textit{proofs of correct encryption}. Property 2 of S-VE definition holds because Lemma \ref{lemma2} is true for all $k$ in the m-Segmentation of $x$. Assuming DDH is hard: The encryption scheme is CPA secure (lemma \ref{lemma1}) and bulletproof is zero knowledge range proof (\cite{bulletproof} Definition 13).

\section{Atomic Cross-Chain Swap Protocol}
\subsection{Background}
Without many exceptions, all blockchains today are using \oded{a} digital signature algorithm that has \oded{an} existing variant for threshold signing. Most dominant are ECDSA and Schnorr based signatures \oded{(including EdDSA)} \cite{boneh17,lindell17,doerner,goldfeder18,lindell18,maxwell,cosi}. In addition BLS signatures show great promise for \oded{the} blockchain use case \cite{boneh18}. A wallet is a user interfacing software that generates and manages private keys, constructs and signs messages (transactions) for the relevant cryptocurrencies. Our wallet design is based on $\{t,n\}$-Thresh-Key-Gen and $\{t,n\}$-Thresh-Sign with minimal amount of participants. We aim to fit it to a simple client-server communication model and therefore take $t = n = 2$.
For the fair exchange protocol we will need a $\{3,3\}$ \oded{threshold signature scheme}.
We assume that multiple cryptocurrencies are supported by this wallet.
\subsection{Roles and Network} We define two roles: \textit{owner} and \textit{provider}. The owner is the end user who owns the funds in the account and holds one secret share of the secret key. The provider is another share holder of the secret key but \oded{does not provide funds to the shared account}. Its role is to provide the additional security in the system aiding and enabling the owner to generate keys and transact in distributed fashion. From network perspective, one provider is connected to many owners which together maintain the provider, for example paying its cost in transaction fees. The provider can run on any machine: from a Trusted Execution Environment (TEE) to a machine operated by an incentivized human operator. Multiple providers can compete for owners. Access to a full node can be done either by both parties locally or outsourced to a number of other parties. \\

\subsection{Setup}
The provider and owner will run $\{2,2\}$-Thresh-Key-Gen. Public keys are used to generate blockchain addresses.
It is assumed that two parties, $P_1$ and $P_2$, that are sharing the same provider (both run $\{2,2\}$-Thresh-Key-Gen with the provider) wish to conduct an exchange between two assets on two different chains. \oded{In details, each party $P_i$ holds funds (i.e. tokens) on the blockchain they wish to swap from, $b_i$, in an input $\{2,2\}$-address we denote $A^{p_i}_{in}$. The amount of funds held by each party should be at least the amount they wish to swap, denoted as $c_i$.\newline
Each party $P_i$ would also have a withdraw address we denote $A^{p_i}_{out}$, on the blockchain they wish to swap to, $b_{3-i}$. This address is not necessarily generated in advance and can be generated later as the protocol progresses, but for sake of simplicity we define it in the setup phase.} Note: we provide a description in terms of addresses but an exact description holds with respect to unspent transaction outputs (UTXOs).\newline
The amounts and maker/taker roles are matched by a public matching engine.\newline

\subsection{The Protocol}
\begin{mdframed}[userdefinedwidth = 12.3cm]
\begin{enumerate}
\item \oded{Party $P_1 (P_2)$ runs \textit{Juggling.KeyGen} to create a decryption/encryption key pair $y_1$ and $Y_1$ ($y_2$ and $Y_2$). The parties then exchange each other's public (encryption) keys $Y_1$ and $Y_2$.}
\item $P_1,P_2,S$ run $\{3,3\}$-Thresh-Key-Gen: The output is public key $pk_1$ and secret shares $x^{p_1}_1,x^{p_2}_1,x^{s}_1$ to each party. We call $Q^i_1$ the \textit{local public key} of party $i$ if $x^i_1$ is the elliptic curve DLog of $Q^i_1$. $pk_1$ is used to derive address $a_1$ in blockchain $b_1$.
\item $P_1,P_2,S$ run $\{3,3\}$-Thresh-Key-Gen again with outputs $pk_2, x^{p_1}_2,x^{p_2}_2,x^{s}_2$ and $Q^i_2$ for the local public keys. $pk_2$ is used to derive address $a_2$ in blockchain $b_2$.
\item $P_1 (P_2)$ runs $\{2,2\}$-Thresh-Sign with the provider to transfer from $A^{p_1}_{in} (A^{p_2}_{in})$ an amount of $c_1 (c_2)$ tokens to $a_1 (a_2)$. The provider $S$ will broadcast the two signed transactions to the two blockchains simultaneously.
\item $P_1$ and $P_2$ run \textit{Juggling.Encryption} with inputs $\{x^{p_1}_1,Q^{p_1}_1,Y_2\}$ and $\{x^{p_2}_2,$ $Q^{p_2}_2,$ $Y_1\}$ respectively. Encryption is done with segmentation enabled: $P_1$ and $P_2$ run \textit{Juggling.Encryption} up to step (d). If one of the previous steps failed - abort. \\
Repeat for $k \in \{1,...,m\}$:
\begin{enumerate}
\item $P_1$ sends $E_k$ (encryption of $[x^{p_1}_1]_k$) and corresponding proof of correct encryption.
\item $P_2$ and $S$ verify proof of correct encryption (aborts if false) and decrypt the segment encryption. $P_2$ and $S$ send back  $E_k'$ (encryption of $[x^{p_2}_2]_k$) and corresponding proof of correct encryption.
\item $P_1$ and $S$ verify proof of correct encryption (abort if false) and decrypt the segment

\end{enumerate}
\item After decryption is done $P_1(P_2)$ and $S$ run \{3,3\}-Thresh-Sign to sign \oded{the withdraw transaction which empties $a_2(a_1)$ into $A^{p_1}_{out}$ $(A^{p_2}_{out})$.\newline The transaction can now be broadcast by either $S$ or $P_1(P_2)$}.
\end{enumerate}
\end{mdframed}

\begin{figure}[htbp]
\centering
  \includegraphics[trim=1.2cm 7cm 1.2cm 0, clip, width=\textwidth]{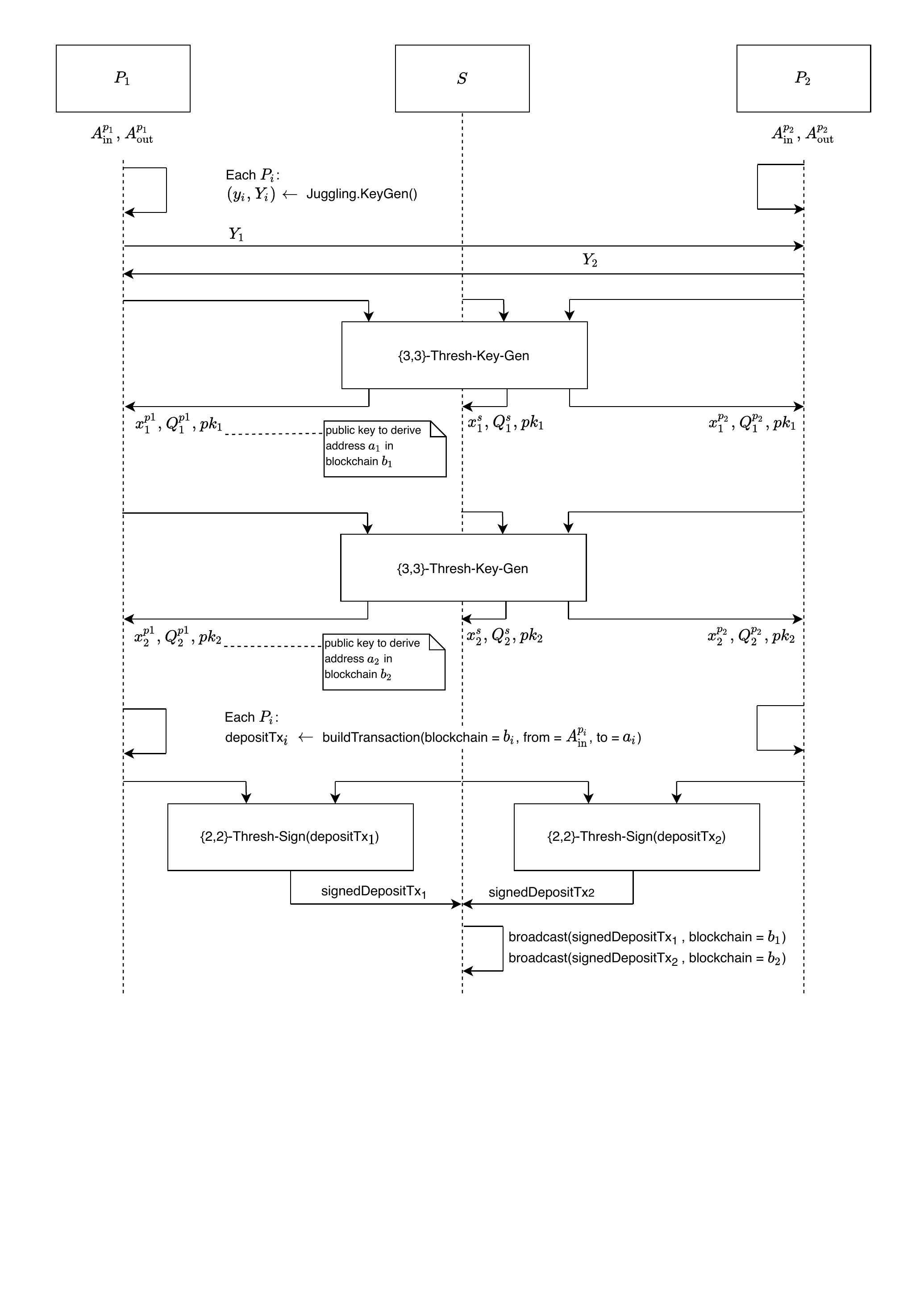}
  \caption{\textbf{The Atomic Swaps protocol sketch - part 1}}
  \label{fig:atomic-swaps-sketch-part1}
\end{figure}

\begin{figure}[htbp]
\centering
  \includegraphics[trim=1.2cm 7cm 1.2cm 0, clip, width=\textwidth]{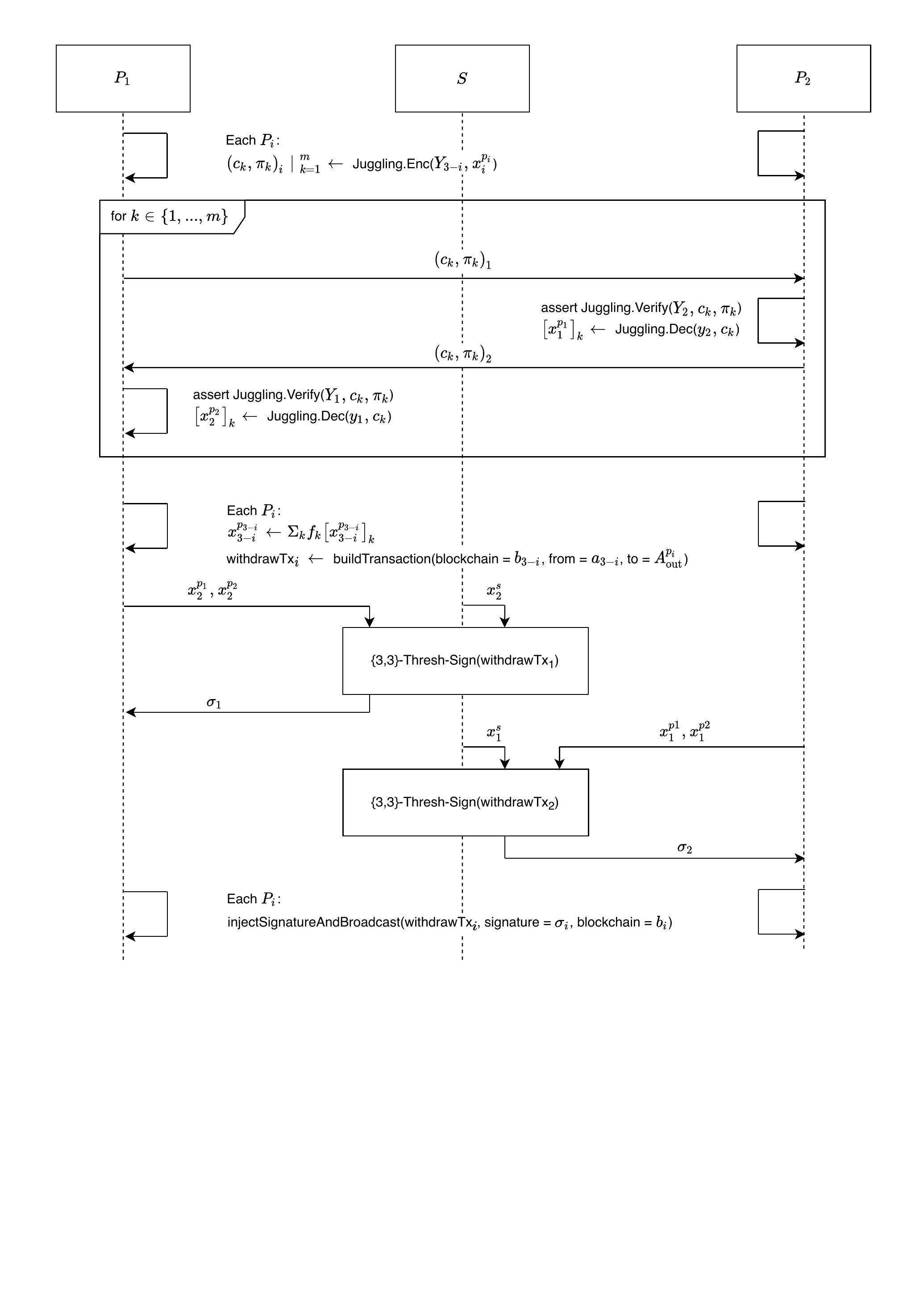}
  \caption{\textbf{The Atomic Swaps protocol sketch - part 2}}
  \label{fig:atomic-swaps-sketch-part2}
\end{figure}

\section{Analysis}
In this section we give informal discussion on security aspects of the protocol. First, we show correctness of the scheme. Second, we show that the protocol guarantees robustness against malicious adversaries and in particular we show that users of the underlying $\{2,2\}$-key management infrastructure do not need to make any additional security assumptions when conducting a swap. Finally, we discuss practical aspects such as cost and privacy. 
\subsection{Use of Juggling} The security of the scheme relies on the security of Juggling scheme: The protocol consists basically of two \oded{interleaving} Juggling protocols\oded{, where the two trading parties are switching the roles of the encryptor and the decryptor}. Key generation is done as part of the creation of $\{3,3\}$-addresses. Encryptor $P_i$ encrypts its secret share of the $\{3,3\}$-public key with $Y_{3-i}$. The segmentation property of the encryption together with the publicly verifiable proofs are taking care of the partial fairness of the exchange: if segment $k$ is false both parties $P_1,P_2$ will either have the same amount of already decrypted segments or one of the parties will have one less decrypted segment. Assuming small segments and that the protocol stopped at a stage where extraction of the rest of the secret share is possible, both parties will need the same resources to get the full secret share. We discuss later the case where one party aborts very early in the protocol such that full extraction is not possible.\\ After all encrypted segments were transferred to the other party and decrypted, the $\{3,3\}$-addresses are \oded{effectively} degenerated into $\{2,2\}$-addresses between the provider $S$ and the counterparty which is the new owner. \oded{The owner can eventually spend from these addresses using the $\{3,3\}$-Thresh-Sign algorithm by contributing its own secret share created at the key generation phase and the counterparty's secret share it has successfully decrypted, together with the provider's secret share.}

\subsection{The Role of the Provider}
\label{provider-role}
The provider $S$ can serve as a PKI because it is a centralized authority used by all owners to derive public keys. This can be strengthened in several ways. For example: assuming $S$ public key is known to all, $S$ can generate a certificate for every $\{2,2\}$-address that it takes part in generating. This certificate can be checked by the participating parties in a swap as well as by an external matching engine.\\
Furthermore, The provider can be decentralized \oded{(i.e. broken into multiple signing parties)} or run on a TEE. Upon detection of a \oded{flawed} segment encryption by one of the parties, $S$ can revoke the party from using the swap service or even from the wallet service.\\
\oded{A provider may be financially incentivized and charge fees for providing the additional security and enabling the fair exchange protocol.\\
The provider is trusted for:
\begin{enumerate}
    \item Availability.
    \item Fair submission of deposit transactions - in step 4 in the protocol the provider is required to submit both deposit transactions which fund $a_1$ and $a_2$, or none in case one of the owners did not send its own transaction.
\end{enumerate}
If assumption (1) is not satisfied, the protocol would not be able to progress and in most scenarios both parties' funds would be kept locked. In particular, if the provider would interact in a $\{3,3\}$-Thresh-Sign with one party but not with the other, then the latter would have its funds locked but the one which was able to get the signature has actually succeeded to complete the trade and got the counterparty's deposit.
If assumption (2) is not satisfied, one party might lock funds by having its deposit transaction submitted by the provider and confirmed, while the other party has aborted and did not stake anything. We should also wrap within this same assumption the requirement that the provider will not collude with a party to \textit{revert} an unconfirmed deposit transaction, which would lead to the same described situation. Since the input addresses created by $\{2,2\}$-Thresh-Key-Gen with the provider, it is able to prevent specifically such attempt.\newline
However, it is important to highlight that in terms of potential outcomes for the dishonest provider: the provider can \textit{freeze} funds but it cannot \textit{steal} funds. The latter means that it cannot extract secret keys nor trick participants to move funds to an arbitrary address of its choice. Also note that violations of the aforementioned trust assumptions can be \textit{publicly verifiable} - an owner may accuse a provider by presenting a transcript of the threshold key generation protocols and optionally the Juggling gradual release protocol, so that an external observer of the relevant blockchains can witness that only one deposit or withdraw transaction has been submitted. That way the provider can be held accountable for denying service from the owner. 
Combined with the economical incentive potential for an honest behavior, these points may give further justification for the trust assumptions.}

\subsection{Adversary Model}
Our goal is to not add additional assumptions on adversarial behavior on top of what we already assume for a $\{2,2\}$-wallet design with an owner and a provider, namely: (1) the provider can stop the service, (2) the provider cannot steal secret keys. The first assumption is dealt outside the scope of this paper: we assume that each party has a recovery method for $\{2,2\}$-addresses, independent of the provider, to reclaim the full secret key. \oded{In our protocol, the provider is assumed to be trusted for the specific points detailed in Section \ref{provider-role}, namely, availability and fair submission of deposit transactions which preserve the aforementioned adversarial behavior limitations (and is not trusted for anything else e.g., for integrity). The owners $P_1,P_2$ can possibly turn malicious at any point.}
In case one of the owners $P_1,P_2$ become malicious, the publicly verifiable nature of the encryption is used to catch the bad actor.
The provider $S$ can play the role of one of the parties. \oded{Without loss of generality, say that $S$ plays the role of (or colludes with) $P_2$. In such a case, note that $P_1$ will not start the Juggling protocol (i.e. the gradual release secret exchange protocol) until it verifies a confirmed transaction transferring the expected amount ($c_2$) into the expected address $a_2$ on the destination blockchain $b_2$. $P_1$ has participated in the creation of this specific address through the $\{3,3\}$-Thresh-Key-Gen algorithm. Therefore, the provider $S$ will be able to get $P_1$'s secret share only if it has funded $a_2$ with sufficient funds which are redeemable by $P_1$. This is equivalent to the honest case where a counterparty $P_2$ doesn't collude with the server and deposits the needed funds.\newline
Lastly, a malicious party may try to withdraw \textit{both} of the deposits from $a_1$ and $a_2$. W.l.o.g. say it is $P_2$. Here, $P_1$ has a secret share $x^{p_1}_2$ it has contributed to the second $\{3,3\}$-Thresh-Key-Gen, which is required for a successful $\{3,3\}$-Thresh-Sign and is not revealed to any other party (even not through the Juggling protocol's gradual release). So even when the provider colludes with $P_2$, they will not be able to withdraw funds from $a_2$ without $P_1$'s cooperation.}

\subsection{Practical Aspects} It is important to note that unlike other ACCS protocols, in this protocol we do not make any assumptions on the blockchains in the trade. This means that in the general case either both parties completed the trade or that both parties did not complete the trade: there is no going back and the funds will stay locked. If the blockchains used have a time lock mechanism it can smoothly be plugged in the protocol: $S$ and the parties will sign funding transactions with time locks such that once they are published if one party decides to go back both parties will get the refund after the time locks are over. \\
\oded{In terms of interoperability, the protocol enables swaps between popular ECDSA blockchains, e.g. Bitcoin and Ethereum, and emerging EdDSA blockchains such as Facebook's Libra, Algorand and Tezos\footnote{Tezos does have support for ECDSA under the curve Secp256k1 as in Bitcoin and Ethereum, but exchanging into an address of a different signing algorithm may not be ideal in terms of wallets and tools support.}. 
This is not possible with existing scriptless ACCS protocols \cite{poelstra2017scriptless,malavolta2019anonymous} that require both blockchains to use one of Schnorr or ECDSA signature schemes that share the same elliptic curve group generated by the same fixed generator.}
The footprint of the protocol on the blockchain is two standard transactions on $b_1$ and two standard transactions on $b_2$. The cost of the protocol is two standard transactions for every party. Using the scheme to swap assets on the same blockchain or using it twice to complete the path $b_1 \rightarrow b_2 \rightarrow b_1$ will result in a mixing solution. 

\section{Implementation}
As a proof of concept we show an atomic swap between Bitcoin and Ethereum. For key management system we  implemented Lindell's \cite{lindell17} $\{2,2\}$ ECDSA key generation and signing protocols (protocols $3.1$ and $3.2$ in \cite{lindell17})  for Elliptic curve Secp256k1. This is the digital signature scheme and elliptic curve used by both Bitcoin and Ethereum. Our code abstracts both the elliptic curve layer and the digital signature layer such that other types of elliptic curves and digital signatures can be plugged in easily. We tested it with curve25519 and library we wrote for multi party Schnorr signatures based on \cite{micali}. \\
We used \oded{$m=32$} segments of \oded{$l=8$} bits for the 256 bit keys. We used brute-force to find the DLogs which is not optimal and in practice one of known protocols that can achieve $2^{l/2}$ time should be used. \\
We implemented the fair exchange protocol for the case where the provider is not allowed to play $P_1$ or $P_2$. This allows to relax the requirement on the of $\{3,3\}$ addresses and use only $\{2,2\}$-Thresh-key-generation and Signing. 
The cryptography layer is written in Rust. \oded{The wrapping communication layer between the parties and their blockchain nodes are written in JavaScript (Node.js)}. It is open source and can be found in \oded{\cite{atomicswaps}}. 
\subsection{Future Work}
We plan to add support for more blockchains: connecting other types of elliptic curves and threshold signature algorithms to the system. For threshold-ECDSA we will switch to a newer protocol \cite{goldfeder18} with support for $\{3,3\}$ addresses. We wish to run experiments to set up optimal system parameters (i.e. optimal bit length $l$) and to measure performance. 

\bibliographystyle{abbrv}
\bibliography{Bibliography}

\begin{appendix}
\section{Security for Auxiliary Proofs}
Here we provide security proofs that the auxiliary proofs from section \ref{aux_proofs} are secure.
\subsection{Proof of Correct Encryption:}
\begin{itemize}
\item \textbf{Completeness:} $z_1G + z_2Y = s_1G + exG + s_2Y + erY = A_1 + A_2 + e(xG+rY)= T + eD$. $z_2G = s_2G +_erG = A_3 + eE$
\item \textbf{Special Soundness:} Given two transcripts $\{T,A_3,e,z_1,z_2\}$ and \\ $\{T,A_3,e',z_1',z_2'\}$ it is easy to extract $r$ from $e,e',z_2,z_2'$ and the $x$ from $e,e',z_1,z_1'$.
\item \textbf{Special HVZK}: Given the statement $G,Y,D,E$ and $e$, choose random $z_2\in Z_q$ and compute $A_3 = z_2G -eE$. Choose random $z_1 \in Z_q$, and compute $T = z_1G + z_2Y - eD$. 
\end{itemize}
\subsection{Proof of Correct Encryption of DLog:}
\begin{itemize}
\item \textbf{Completeness:} $z_1G=s_1G + exG = A_1 + eQ$, $z_2G = s_2G + erG = A_3 + eE$, $z_2Y = s_2Y + erY = A_2 + eQ+erY -eQ = A_2 + e(D-Q)$
\item \textbf{Special Soundness:} Given two transcripts $\{A_1,A_2,A_3,e,z_1,z_2\}$ and \\ $\{A_1,A_2,A_3,e',z_1',z_2'\}$ it is easy to extract $r$ from $e,e',z_2,z_2'$ and the $x$ from $e,e',z_1,z_1'$.
\item \textbf{Special HVZK}: Given the statement $G,Y,Q,D,E$ and $e$, choose random $z_2\in Z_q$ and compute $A_3 = z_2G -eE$, $A_2 = z_2Y -e(D-Q)$. Choose random $z_1 \in Z_q$, and compute $A_1 = z_1G - eQ$. 
\end{itemize}
\end{appendix}
\end{document}